\documentclass[useAMS,usenatbib,letterpaper]{mn2e}
\usepackage{graphicx}

\begin{document}


\title[GRB 060912A and the long short divide]
{A case of mistaken identity? GRB 060912A and the
nature of the long -- short GRB divide}

\author[A.J.~Levan et al.]{A. J.~Levan$^{1}$\thanks{email: a.j.levan@warwick.ac.uk, 
Based on observations made with ESO telescopes at the Paranal Observatory under 
programme ID 077.D-0691},
P.  Jakobsson$^{2}$, C. Hurkett$^{3}$, N.R. Tanvir$^{3}$, 
J. Gorosabel$^{4}$, \and P. Vreeswijk$^{5}$,  E. Rol$^{3}$, R. Chapman$^{2}$,
N. Gehrels$^{6}$, 
P.T. O'Brien$^{3}$, J.P. Osborne$^{3}$,  \and
R.S. Priddey$^{2}$, C. Kouveliotou$^{7}$, R. Starling$^{3}$, 
D. Vanden Berk$^{8}$, K. Wiersema$^{9}$\\
$^1$Department of Physics, University of Warwick, Coventry, CV4 7AL, UK \\
$^{2}$Centre for Astrophysics Research, University of
Hertfordshire, College Lane, Hatfield AL10 9AB, UK.\\
$^{3}$Department of Physics and Astronomy, University of Leicester,
Leicester, LE1~7RH, UK. \\
$^{4}$ Instituto de Astrof\'{\i}sica de Andaluc\'{\i}a (IAA-CSIC),
     Apartado de Correos, 3.004, E-18.080 Granada, Spain.\\
     $^{5}$ European Southern Observatory, Alonso de C—rdova 3107, Casilla 19001, Vitacura, Santiago, Chile \\
     $^{6}$ NASA Goddard Space Flight Center, Greenbelt, MD20771, USA \\
     $^{7}$ NASA Marshall Space Flight Centre, NSSTC, SD-50, 320 Sparkman Drive, Huntsville, AL 35805, USA \\
     $^{8}$ Department of Astronomy and Astrophysics, The Pennsylvania State University, 525 Davey Laboratory, University Park, PA 16802 \\
$^{9}$ Astronomical Institute `Anton Pannekoek', University of Amsterdam, Kruislaan 403, 1098 SJ Amsterdam, The Netherlands
}

\date{Accepted 2007 April 18. Received 2007 April 15; in original form 2007 March 7}


\maketitle

\label{firstpage}

\begin{abstract}
 We investigate the origin of the GRB 060912A, which has observational
  properties that make its classification as either a long or short burst
  ambiguous.  Short duration GRBs (SGRBs) are thought to have typically
  lower energies than  long duration bursts, can be found in galaxies with
  populations of all ages and are likely to originate from different progenitors
  to the long duration bursts. However, it has become clear that duration alone
  is insufficient to make a distinction between the two populations in
  many cases, leading
  to a desire to find additional discriminators of burst type.
  GRB 060912A had a duration of 6 s and occurred
  only $\sim$ 10\arcsec~from a bright, low redshift ($z=0.0936$) elliptical galaxy,
  suggesting that this may have been the host, which would
  favour it being a short-burst. 
  However, our  deep
  optical imaging and spectroscopy of the location of GRB 060912A
  using the VLT
  shows that GRB 060912A more likely
  originates in a distant star forming galaxy at
  $z=0.937$, and is most likely a long burst.
  This demonstrates the risk
  in identifying bright, nearby galaxies as the hosts of given GRBs
  without further supporting evidence. 
  Further,
  it implies that, in the absence of secure identifications, ``host"
  type, or more broadly discriminators which rely on galaxy redshifts,
  may not be good indicators of the true nature of any given GRB.
\end{abstract}


\section{Introduction}
Observations since the launch of Swift have finally begun to shed
light on the nature of short duration GRBs (SGRBs - Kouveliotou et al. 1993).
These observations demonstrate
their apparent origin in populations of all ages (Gehrels et al. 2005; Hjorth
et al. 2005a; Fox et al. 2005; Berger et al. 2005; Bloom et al. 2006) and, at lower
redshift on average than the long duration bursts (Jakobsson et al. 2006a), 
now known to originate in
stellar core collapse (Hjorth et al. 2003; 2005b; Stanek et al. 2003). 
However, these observations have
also emphasised the key issue of the distinction between 
long and short duration GRBs, as the
two populations have significant overlap in many of
their observed properties. 
Thus the task of accurately identifying a given burst as 
belonging to the long or short population is of 
particular importance.

Scientifically, the primary motivation for distinguishing between
short and long GRBs (and indeed the true physical difference
between the two subclasses of event) is the putative association
of each class with a different mechanism for the production of the GRB.
The difficulties in making this distinction are perhaps most strikingly 
illustrated by the low-redshift bursts GRB 060505 and 060614 which, while both
exhibiting durations of $>$2s, 
were not associated with bright supernovae, 
and may therefore represent another progenitor type
(Fynbo et al. 2006; Gal-Yam et al. 2006; Della Valle et al. 2006; 
Gehrels et al. 2006). 

Various possible criteria have been suggested for distinguishing what
is truly a short population GRB (although perhaps the
broader question is to identify which
bursts may not be due to collapsars), these have been addressed by
Donaghy et al. (2006). Of crucial importance are:

\begin{itemize}

\item
{\em The sensitivity of the instrument making the detection and the 
energy range in which it operates}. For example, {\it Swift}/BAT operates
primarily in the 15-150 keV range, which is softer than the 50-350 keV
range where BATSE was most sensitive. As it is known that GRB emission
lasts longer at lower energies (and also at higher 
sensitivities where the decay of the burst can be followed for longer)
this needs to be taken into account. For example, the 
GRBs 050724 ($t_{90} =3$ s) and 050911 ($t_{90}=16$ s) would both have
been classified as short duration GRBs when viewed by BATSE
(Barthelmy et al. 2005; Page et al. 2006).

\item
{\em The spectral properties of the prompt emission}. Short bursts
are (on average) spectrally harder than long bursts 
(Kouveliotou et al. 1993), thus hard 
$\gamma$-ray emission is a good diagnostic. Additionally 
short bursts show light curves which correlate well in all
bands, while in long bursts the softer emission "lags" behind
the harder emission (e.g. Norris \& Bonnell 2006).

\item
{\em The properties of the host galaxy}. Long bursts occur
primarily in sub-luminous blue star forming galaxies
(e.g. Le Floc'h et al. 2003; Christensen et al. 2004;
Fruchter et al. 2006). In contrast the suggested host galaxies
of several short GRBs are bright elliptical galaxies
at moderate redshift (e.g. GRBs 050509B and 050724 Gehrels
et al. 2005; Bloom et al. 2006a; Berger et al. 2005), although
some host galaxies are star forming (e.g. Fox et al 2005),
or at much higher redshift (Levan et al. 2006b; Berger et al. 2006a).
This has led to the suggestion that early-type host galaxies
are a sufficient but not necessary indicator that
a burst is of the short class.

\item
Based principally on GRB 050509B (Gehrels et al. 2005; Pedersen 
et al 2005) it has also been suggested that SGRBs may occur in
greater numbers in galaxy clusters. Although further searches
for clusters associated with SGRBs have found few other examples
(e.g. Berger et al. 2006b), implying that (at the very least) the absence
of a cluster cannot rule out an SGRB origin. Equally,
relatively few searches for associated clusters have been conducted for
LGRBs (see Levan et al. 2006c), and so the comparative 
properties remain poorly understood.

\item
{\em The energy of the burst, and presence of a supernova}.
The long GRB population is now moderately well studied, 
with a large sample of redshifts and isotropic $\gamma$-ray energy releases. 
Equally, until recently, essentially all well studied LGRBs at $z < 1$ 
showed late time signatures plausibly associated with supernovae
(e.g. Zeh et al. 2004). 
Deviations from these properties are obvious causes for interest,
and may indicate a different population of bursts.

\end{itemize}

Here we consider the case of the intermediate duration
($t_{90} = 5.98 \pm 0.07$s) burst GRB 060912A.
The proximity of this burst to a bright, elliptical galaxy at $z=0.0936$
led to speculation that this was the host (Berger 2006). 
Here we present deep optical 
imaging and spectroscopy which was able to pinpoint
the optical afterglow on the sky and demonstrate that the burst
most likely originated at markedly higher redshift ($z=0.937$). 
This illustrates the dangers of making such associations and may have
interesting consequences for our view of short burst host identifications
to date, most notably those of GRB 050509B (Gehrels et al. 2005; Bloom et al. 2006)
and 060502B (Bloom et al. 2007).

\begin{figure}
\begin{center}
\centerline{
\resizebox{8truecm}{!}{\includegraphics[angle=270]{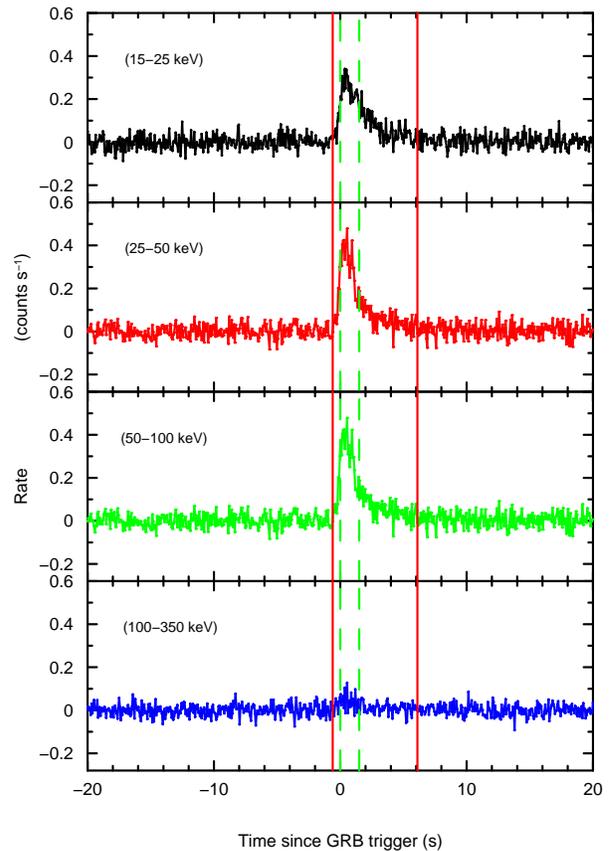}}}
\end{center}
\caption{The BAT lightcurve of GRB 060912A in 4 energy bands (from top
to bottom 15-25 keV, 
25-50 keV, 50-100 keV, 100-350 keV). The $t_{50}$ (dashed lines) and $t_{90}$ 
(solid lines) durations  are also
shown.}
\label{f1}
\end{figure}

\section{Observations}
GRB 060912A was detected by {\it Swift} at  13:55:54 UT on 12 September 2006
(Hurkett et al. 2006a). It exhibited a bright X-ray and optical
counterpart at a location of RA= 00$^{h}$ 21$^{m}$ 08.16$^{s}$ Dec =
+20$^{d}$ 58\arcmin~17.8\arcsec (Hurkett et al. 2006b). 
The burst was also detected by
the UVOT in all but the bluest (UVW2) band (Brown \& Hurkett 2006).
The BAT lightcurve in 4 bands is shown in Figure 1, with the $t_{50}$ and
$t_{90}$ durations marked, we derive $t_{90} = 5.98 \pm 0.07$s and
$t_{50}  = 1.98 \pm 0.05$s. These values are somewhat longer
than the typical values for BATSE short bursts ($t_{90} <2$s, $t_{50} < 1$s; 
Kouveliotou et al. 1993), although recent
observations of short bursts and a reanalysis of the duration
distributions from BATSE bursts suggest that short population
bursts can have $t_{90} > 2$s (Donaghy et al. 2006; 
Zhang et al. 2007). The burst was also detected by
Konus-Wind in the 20-2000 keV energy range, and exhibited a duration 
of $\sim 8$ s (Golenetskii et al. 2006). The measured
photon indicies are $1.74 \pm 0.09$ for
the BAT over the 15-150 keV range (Parsons et al. 2005) and $1.94 \pm 0.2$ from
the Konus-Wind data in the 20-2000 keV range.

The location of the burst lies offset approximately 11.5\arcsec~ from a bright
elliptical galaxy with $K=13.2$. This galaxy was found to lie
at $z=0.0936$ and thus has $M_K \sim -25$, only marginally fainter
than the cD galaxy found close to the location
of GRB 050509B which had $M_K \sim -25.5$ (Gehrels et al. 2005). 

We observed the location of GRB 060912A using the VLT and FORS1 on
14$^{th}$ September 2006.  At the location of the optical afterglow (Hurkett et 
al. 2006a) we found a clearly extended source which we identified
as the host galaxy of GRB 060912A (Levan et al. 2006a - see section 3.3 for
further discussion on the allocation of the host galaxy).
A image of the host galaxy is shown in Figure 2, while an enlarged field is 
shown in Figure 3.
 We subsequently obtained
a spectrum of the host galaxy using the VLT and FORS2 on 21 September 2006. 
The
spectrum exhibits a single, very strong emission line at 7219\AA,
which we identified as  [\mbox{O\,{\sc ii}}] (3727\AA) at a redshift of $z=0.937$ 
(Jakobsson et al. 2006b);
we also detected lines of   [\mbox{Ne\,{\sc iii}}] and H$\gamma$ at the same
redshift, with a confidence of $8 \sigma$ and $4 \sigma$ respectively.
An alternative explanation of the strong line is that it is due to H$\alpha$
at z=0.0999, however in this case additional weaker emission lines due to [NeIII]
and H$\gamma$ could not be explained. Additionally, given the strength of the 
emission line, 
the absence of bluer
emission lines due to  [\mbox{O\,{\sc ii}}], [\mbox{O\,{\sc iii}}] and 
H$\beta$ would be surprising.
The emission lines from  [\mbox{O\,{\sc ii}}] and 
[\mbox{Ne\,{\sc iii}}] seen from the host of GRB 060912A 
are typical of the host galaxies of long duration GRBs (e.g. Bloom et al. 1998;
Vreeswijk et al. 2001,2006). The equivalent width of the [\mbox{O\,{\sc ii}}]  line is found to 
be 130\AA, amongst the largest EW seen in galaxies of comparable
magnitude (e.g. Glazebrook et al. 1994). The
total emission line flux is $F = 6.6 \times 10^{-17}$ ergs s$^{-1}$ cm$^{-2}$ and 
converting this to a luminosity, and subsequently a star formation rate
using the method of Kennicutt (1998) implies 
SFR= 4 M$_{\odot}$ yr$^{-1}$ (formally a lower limit on
the star formation since no extinction correction is possible)\footnote{
Calculations have been performed assuming a $\Lambda$CDM
cosmology with $\Omega_M = 0.27$, $\Omega_{\Lambda}=0.73$ and
$H_0 =71$ km s$^{-1}$ Mpc$^{-1}$}. 
Although there remained some afterglow
contamination at the time of our spectroscopic observations, by
subtracting a point source from the location of the afterglow we obtain
an estimate of the host galaxy magnitude to be R=22.0 $\pm$ 0.5
(the significant uncertainly stems from the afterglow subtraction).
 This corresponds to an absolute magnitude
of $M_R \sim -21$, and is amongst the brightest GRB host galaxies 
(Fruchter et al. 2006). The offset of the GRB from the brightest region
of the host galaxy is $\sim 0.3 \pm 0.1$\arcsec, corresponding
to $2.3 \pm 0.8$ kpc, again typical of long GRBs (Bloom et al. 2002). 

Additionally, when obtaining spectra of the host galaxy we aligned
the slit such that two additional bright galaxies lay across it. These
galaxies lie at
 RA = 00$^{h}$ 21$^{m}$ 05.5$^{s}$, 
Dec = +20$^{d}$ 58\arcmin 18.1\arcsec, and 
 RA = 00$^{h}$ 21$^{m}$ 00.5$^{s}$, 
Dec = +20$^{d}$ 58\arcmin 18.0\arcsec (see Figure 3).
We identify these to be at $z=0.0936$ and $z=0.0977$, based on
absorption lines from Ca H\&K, H$\beta$, H$\delta$, 
[\mbox{Mg\,{\sc i}}], [\mbox{Na\,{\sc i}}],  and the g-band.
A previously catalogued source at RA = 00$^{h}$ 21$^{m}$ 19.1$^{s}$, 
Dec = +21$^{d}$ 00\arcmin 25.00\arcsec~ has a redshift of 
$z=0.0945$, making a 
total of at least 4 galaxies within 2\arcmin~ with similar redshifts, suggesting
a foreground overdensity.

\begin{figure*}
\begin{center}
\centerline{
\resizebox{17truecm}{!}{\includegraphics{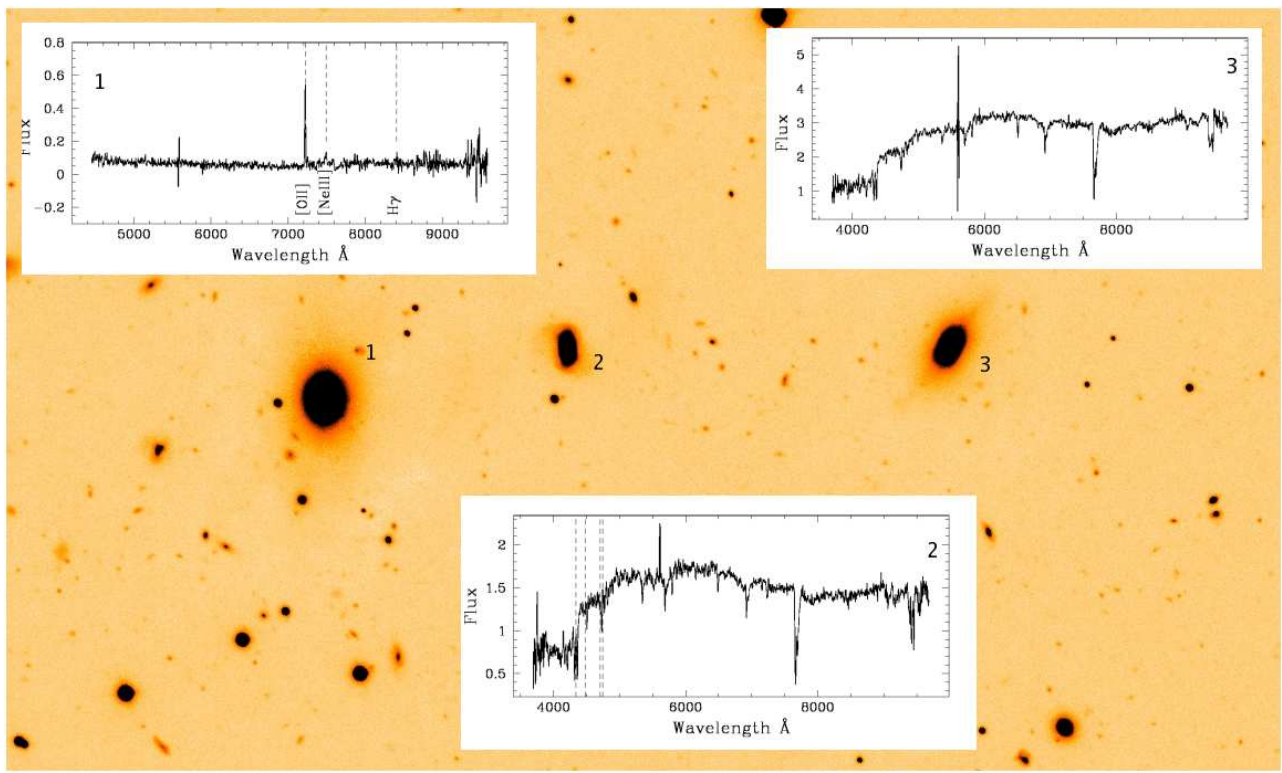}}}
\end{center}
\caption{The region around GRB 060912A. The host galaxy is marked to the left of
the 1. The individual galaxy spectra for the host galaxy and for two other cluster
members are plotted as marked, the host galaxy is just to the left 
of the label. The flux units on the y-axis are $10^{-17}$ 
ergs s$^{-1}$
cm$^{-2}$ \AA$^{-1}$. }
\label{f1}
\end{figure*}

In order to search for putative X-ray emission from the cluster
we summed the available XRT data
taken in photon counting (PC) mode and extracted data in an annulus with inner
and outer radii of 25-250 arcseconds. This allowed us to exclude the region containing 
significant flux from the afterglow. We additionally removed
any other discrete sources contained within the annulus. No statistically significant excess emission
is found within this annulus, and we place a limit of
$F_X < 1.3 \times 10^{-14}$ ergs s$^{-1}$ cm$^{-2}$ (0.3-10 keV) and a corresponding limit
on the X-ray luminosity of the putative cluster of $L_X < 2.8 \times 10^{41} $ ergs s$^{-1}$
(again in the 0.3-10 keV range),
a factor of 10 less luminous that the cluster suggested to be associated with
GRB 050911 (Berger et al. 2006b), this suggests that the observed overdensity
of galaxies is due to a smaller group or cluster.

\section{Discussion}

GRB 060912A had a duration of $t_{90} \approx 6$~s, which is 
near to the $\approx 5$~s duration
which Donaghy et al. (2006) 
find as the point of roughly equal probability of a given burst
lying in either the long or short duration class
(strictly this is an energy- and therefore instrument-dependent
statement, and in this case appropriate for the harder
response of BATSE). 
In such circumstances
it is clearly
necessary to rely on additional information to distinguish between
the two populations.  Previously it has been suggested that host type, 
total energy release and the presence of a supernova component can
be useful in making this distinction (Donaghy et al. 2006), we investigate
each of these in turn.

\subsection{Nearby galaxies as an indicator of burst type}

Host galaxy type is, on first sight an obvious means of distinguishing
between long and short duration GRBs, since some short
bursts are known to originate from elliptical 
galaxies (Berger et al. 2005), while long GRBs require current
star formation. However, in many cases the level 
of significance of the association is rather low, at the
2-4 $\sigma$ level (e.g. GRB 050509B: Gehrels et al. 2005; Bloom
et al. 2006a; GRB 060502B: Bloom et al. 2007), thereby making
the case for assigning a given burst to a host galaxy weaker, and
significantly affecting the use of these galaxies as a proxy for burst type. A common problem
for short burst afterglows is that they are often 
only detected due to the prompt {\it Swift} XRT observations. 
In these cases their locations on the sky are only accurate to 
$\sim 5$\arcsec. Although cross correlating with other sources
can reduce this uncertainty to 2-3\arcsec (e.g. Butler 2007) this
still does not allow for the unambiguous identification of
host galaxies that is possible via sub-arcsecond positions
(Bloom, Djorgovski \& Kulkarni 2002; Fruchter et al. 2006). This
problem may be especially prevalent for S-GRBs, which, if due to
compact object mergers may take place well away from the body 
of the host galaxy.
The issue is well illustrated by Figure 3 which shows the XRT error
circles for both GRB 060912A and GRB 050509B, which are both located
close to bright, low-redshift ellipticals, but as we have seen, in
the former case the association is a chance alignment.

\begin{figure*}
\begin{center}
\centerline{
\resizebox{15truecm}{!}{\includegraphics{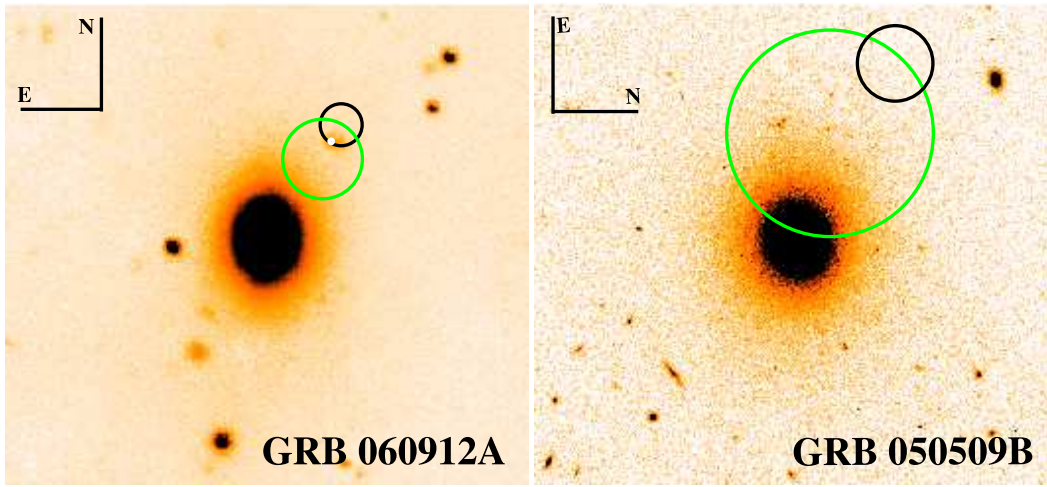}}}
\end{center}
\caption{GRB 060912A and GRB 050509B. In each case the X-ray locations are marked,
both the larger Swift-team XRT error circles and the smaller
revised error circles of Butler (2007) in black. In the case of GRB 060912A the optical
location is also shown in white.
The image of GRB 060912A was taken on 14 September 2006 and shows some 
optical afterglow contribution on the host galaxy within the XRT error box.
In the case of 050509B only an X-ray location was known, and, as can 
be seen in the deep
HST observation in the right hand panel a large number of optical 
sources are visible within
the error box, however none can be firmly associated with GRB 050509B. 
For GRB 060912A the
precise optical position enabled the burst to be located on the host 
galaxy (visible within the
XRT error box), which was subsequently shown to lie at $z=0.937$. Both 
panels are plotted at the same physical scale, 1\arcmin on a side}
\label{f1}
\end{figure*}

\begin{table}
\begin{center}
\begin{tabular}{llll}
\hline
Galaxy & Error (\arcsec)$^{a}$ & $r_i$ (\arcsec) &  P$_{chance}$ \\
\hline
G1  &  3.6$^{b}$ & 12.1 & 0.007 \\
H  &  3.6$^{b}$ & 12 & 0.387 \\
G1 & 1.9$^{c}$ & 13.7 & 0.009 \\
H  &  1.9$^{c}$ & 5.7 & 0.036 \\
G1 & 0.1$^{d}$ & 12.1 & 0.007 \\
H & 0.1$^{d}$ &  0.8 &0.005 \\
\hline
050509B & 9.3$^{b}$ & $\sim$11& $\sim 0.017$\\ 
050509B & 3.4$^{c}$ & $\sim$18 & $\sim 0.027$\\
060502B & 5.4$^{b}$  & $\sim$17.5 & $\sim 0.074$ \\
060502B & 3.7$^{c}$   & $\sim $17.5 & $\sim$ 0.074\\
\hline
\end{tabular}
\end{center}
\caption{Probabilities of chance alignment of GRB 060912A with 
both the nearby elliptical 
galaxy (G1) and the background (assumed host) galaxy at $z=0.937$ (H).
 This is tabulated as a function of 
error box radius, demonstrating the 
importance of obtaining subarcsecond positions
for afterglows, since high redshift galaxies are markedly smaller than an XRT error
circle, which may contain several faint galaxies. The different probabilities 
correspond to different values of $r_i$ which are relevant under the differing
error box sizes and putative host assignments and are described in section 3.1.1. 
The different error radii correspond to i) the optical localisation , ii) The refined
XRT team analysis and iii) The X-ray position of Butler (2007) refined based
on cross-matched astrometry. In each case, following Bloom et al. (2002) we
use the largest value for $r_i$ based on the different means of its estimation 
described in section 3.1.1. For comparison and consistency the values for GRBs 050509B
and 060502B are also calculated using the same approach, these numbers
differ somewhat from those reported previously (e.g. Gehrels et al. 2005; 
Bloom et al. 2006) demonstrating the plausible range depending
on the approach taken. $^{a}$ Errors are given
at the 90\% confidence level. $^b$ Error is given based on the refined analysis
of the XRT data alone. $^c$ Error is determined by cross correlating the X-ray
locations with other optical sources in the field, and is taken from Butler (2007).
$^d$ Error is determined largely from the optical afterglow.}
\label{default}
\end{table}%

\subsection{Energy and supernovae}
Two further diagnostics suggested to distinguish between short and long GRBs
are i) the presence of a supernova and ii) the energy of the burst. It has been
suggested that SGRBs have, in some cases, much lower energy than long bursts. While
no SGRB afterglow has exhibited any supernova signature, which are common in
LGRBs.
However,
these two scenarios are also crucially dependent on secure redshifts, from 
either host galaxies, or preferably absorption lines. For example
GRB 060912A would have an energy of $E_{iso} \sim 3 \times 10^{51}$ ergs, 
at $z=0.937$, typical of long GRBs (e.g. Bloom et al. 2003), while at 
$z=0.0936$ the energy release would have been 
only $E_{iso} \sim 3 \times 10^{49}$ ergs, more typical of the SGRB population
(e.g. GRB 050724 Barthelmy et al. 2005; GRB 050709 Fox et al. 2005; Villasenor et al. 2005).
Similarly, at $z=0.0936$ any supernova would have been easily visible, 
with $R\sim 19$. However,
at $z=0.937$ its peak magnitude would have been $R \sim 25$, beyond the limits
of most observations.

Therefore,
in the absence of a secure redshift the true energy of the burst, and the
expected properties of any supernovae are not assured, and cannot be
used to make strong constraints on which population it belongs to
(this is illustrated in Figure 4). 

These two constraints are further complicated by the fact that some
S-GRBs apparently originate from much higher redshift, with correspondingly
larger energies (e.g. GRB 050813, GRB 060121 and GRB 060313;
Ferrero et al. 2006; Levan et al. 2006b; Berger et al. 2006a; Hjorth et al. 
in preparation) and the absence of supernova signatures in
the apparently long duration GRBs 060505 and 060614 (Fynbo et al. 2006; Gal-Yam et al. 2006).
Both of these factors may represent further
interesting evidence of the overlap between short and long duration bursts.
For example, it is possible that 
GRB 060505 and GRB 060614 originated from 
the same progenitors as short GRBs (e.g. Gehrels et al. 2006; Ofek et al. 2007), but with
higher energy (Zhang et al. 2007). Alternatively these bursts might still be related to
stellar core collapse without the release of sufficient radioactive material
to create an observable supernova (e.g. all of the material has fallen 
directly onto the nascent black hole, Fryer et al. 2006).

\begin{figure}
\begin{center}
\centerline{
\resizebox{9truecm}{!}{\includegraphics[angle=270]{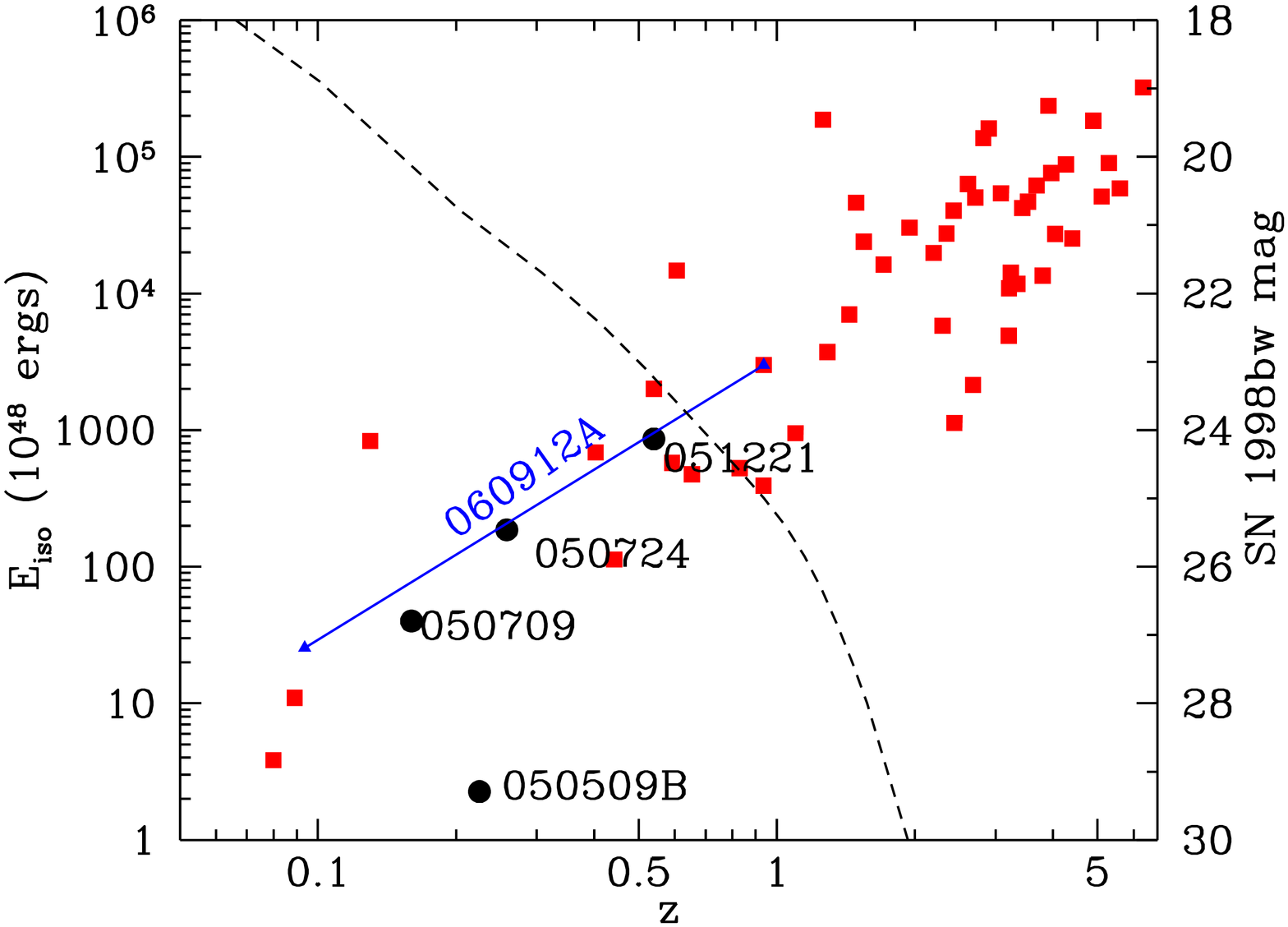}}}
\end{center}
\caption{
The redshift -- energy -- supernova degeneracy. Showing the energies 
(15-150 keV) and redshifts for
{\it Swift} long and short bursts 
(long bursts squares, short burst circles). Also,
showing the energy of GRB 060912A at its different possible redshifts. Any burst with
unknown redshift will essentially move 
along a line of equal gradient. Therefore, as can
be seen moving short bursts with low significance associations with nearby galaxies
out to higher redshifts will result in energies comparable to the long duration bursts. Indeed
several short bursts suggested to be at higher redshift
(e.g. GRB 050813 (Ferrero et al. 2006), GRB 060121 (Levan et al. 2006b) and 
GRB 060313 (Hjorth et al in prep)) may already lie in this region. Additionally shown 
on the right hand axis and with the dashed line is the approximate R-magnitude of a supernova
similar to SN~1998bw at maximum as it evolves with redshift. Beyond $z\sim 1.2$ it
becomes essentially undetectable from the ground. Therefore, in the absence of a clear
redshift identification neither energy, nor the absence of detectable supernova
can place strong constraints on the nature of the burst.}
\label{f1}
\end{figure}

\subsection{Was GRB 060912A long or short population burst?}

We conclude above that the likely host galaxy of GRB 060912A is
a starforming galaxy at $z=0.937$. This implies that the presence
of a nearby elliptical is a chance alignment. However, the reverse
is also possible (i.e. the chance alignment may actually be with the $z=0.937$ galaxy).
To ascertain the likelihoods for this we determine the probability
of a random association as a function of
X-ray error box size, host galaxy magnitude and
host galaxy size.  Following Bloom et al. (2002)
we define a probability of a given GRB  and galaxy being a chance 
association to be:

\begin{equation}
P_{i,ch} = 1 - \exp{(-\eta_i)},
\end{equation}

where

\begin{equation}
\eta_{i} = \pi r_i^2 \sigma(\leq m_i),
\end{equation} 

and

\begin{eqnarray}
\sigma(\leq m_i) = { 1 \over 3600^2 \times 0.334 log_e 10} \nonumber \\  
\times 10^{0.334(m_i - 22.963) + 4.320}
\mathrm{galaxies/arcsec}^2.
\end{eqnarray}

This is based on the R-band number counts from Hogg  et al. (1997),
where $m_i$ is the magnitude of the galaxy in question.
Here $r_i$ is the effective radius of a circle on the sky in which
a given galaxy is found, and is a measure
of angle subtended by the galaxy or (in the case of larger error
localisations) by the errorbox of the GRB. Clearly this depends on the localisation
confidence of the GRB, with 
$r_i = 2 r_h$ (where $r_h$ is the half light radius = 3\arcsec\
for the elliptical and 0.4\arcsec\ for the starforming galaxy)
being suitable for most
long GRBs which lie within the optical light of their hosts, 
$r_i = 3 \sigma_i$ (where $\sigma_i$ is the 1$\sigma$ positional error) 
for large error boxes and $r_i = (4r_h^2 + R_0^2)^{0.5}$
for cases where the burst location lies well 
offset from the optical light of the galaxy
(here $R_0$ is the offset of the burst position from the centre of the putative host). 
This approach enables us to consider {\em both} the uncertainty in the GRB position (which
can be large in the case of XRT locations) and the physical size of the host
galaxy in question (e.g. nearby galaxies are brighter, and so therefore rare, 
but they also have larger angular sizes, 
increasing the probability of a chance alignment). 
The results of using this approach for both galaxies are shown in 
Table 1, which demonstrates the importance of small localisations, 
especially in the case of faint background galaxies. 

The probability of chance alignments with the nearby
elliptical galaxy and with a background galaxy are broadly equivalent
(and indeed can show some variation based on bands used and 
alternative means of estimating $r_i$). One may then wonder why
the higher redshift alternative is favoured. This can be understood in
terms of our prior knowledge of the properties of the host galaxies 
of both long and short duration GRBs. Only a small sample of 
SGRBs have been linked to their hosts with high confidence, with
no cases of absorption redshifts yet reported. So, while the suggestion
that short bursts lie preferentially in older (elliptical) galaxies is 
certainly plausible it has not been demonstrated with high confidence, and
the relative fraction found in such galaxies in poorly constrained
(see e.g. Zheng \& Ramirez-Ruiz 2006). 
In contrast LGRB hosts have been well studied for almost a decade
and large samples now exist (e.g. Le Floc'h et al. 2003; Christensen et al. 2004;
Fruchter et al. 2006). We know that these galaxies are usually
very blue, with
a range of irregular and compact
morphologies (Conselice et al. 2005; Wainwright et al. 2006), and
frequently show strong emission lines (e.g. Vreeswijk et al. 2001).  
We 
are therefore able to make stronger statements about the expected host galaxies of long bursts
than short bursts. Indeed, assuming that the burst was long we would have broadly
predicted {\em a priori} the properties of the $z=0.937$ galaxy as its host. In
other words the probability of the high redshift scenario assuming that the 
burst is long is much higher than the probability of the low redshift scenario
assuming the burst is short (in Bayesian terms $P(H|L)$, where $H$ is high redshift
and $L$ is long burst, is much higher than 
($P(E|S)$ - where $E$ is elliptical and $S$ is short burst). 
An alternative approach would be to re-calculate the probabilities 
of chance alignments using, say, only galaxies which exhibit bright 
emission lines, although we do not do this here it is obvious that such
a cut would significantly reduce the chance of random association with
the $z=0.937$ galaxy, and therefore we believe that our identification of it
as the host of GRB~060912A is justified.

Although we argue above that the high-z origin of GRB 060912A 
is most likely, this discussion illustrates the difficulty of 
definitively 
linking some bursts to their hosts based only on proximity on 
the sky. In particular, had the host of GRB060912A been 
even fainter (as many long-burst hosts are), then it is possible 
that it would not have been discovered and the low-z scenario 
would have been favoured. All this highlights the importance 
of firmly linking
a GRB to
its host galaxy whenever possible, by ensuring that immediate deep
observations are pursued to locate an afterglow, and {\em critically}
also obtaining absorption redshifts for short bursts, which, due to
the general faintness of the afterglows have so far not been
obtained. In some cases it may be possible to constrain redshift
photometrically, however, this is only true for cases where the
redshift is likely to be high enough that the Lyman-$\alpha$ break
passes through one of the UVOT filters (or the optical). In the case
of GRB 060912A (or in other bursts where hosts suggest $z < 1.5$) it
is not possible to use photometric information to greatly constrain
the redshift, or to determine between alternative possibilities.

\subsection{Long -- short overlap: Difficulties in defining a divide}
The difficulties in deciding if a given burst belongs to the long or
short GRB population have been one of the drivers 
for searching for additional constraints on the nature of the burst. However,
as the sample of short GRBs remains small it is very dangerous to propagate the
properties of the small number of bursts seen to date to the larger population,
especially when several of the properties considered have been established
only with marginal significance within the small population. We have discussed 
above the degeneracy that unknown redshift causes when considering the energy
of the burst, the lack of supernova emission or even the nature of the host galaxy. 
However, these degeneracies also affect the prompt emission. Moving a
given burst to higher redshift has two crucial effects. The first is that the burst
duration is time dilated by increasing redshift, while its spectrum is also 
softened (indeed high redshift was initially postulated as an origin for the
very soft emission seen from X-ray Flashes (XRFs e.g. Heise et al. 2001)). The essence of this
problem is therefore simple; {\em redshift can change the emission 
properties of short -- hard bursts in to 
long -- soft bursts}, further complicating any attempts to derive
firm distinctions based on their observed prompt properties. 

\begin{figure}
\begin{center}
\centerline{
\resizebox{9truecm}{!}{\includegraphics{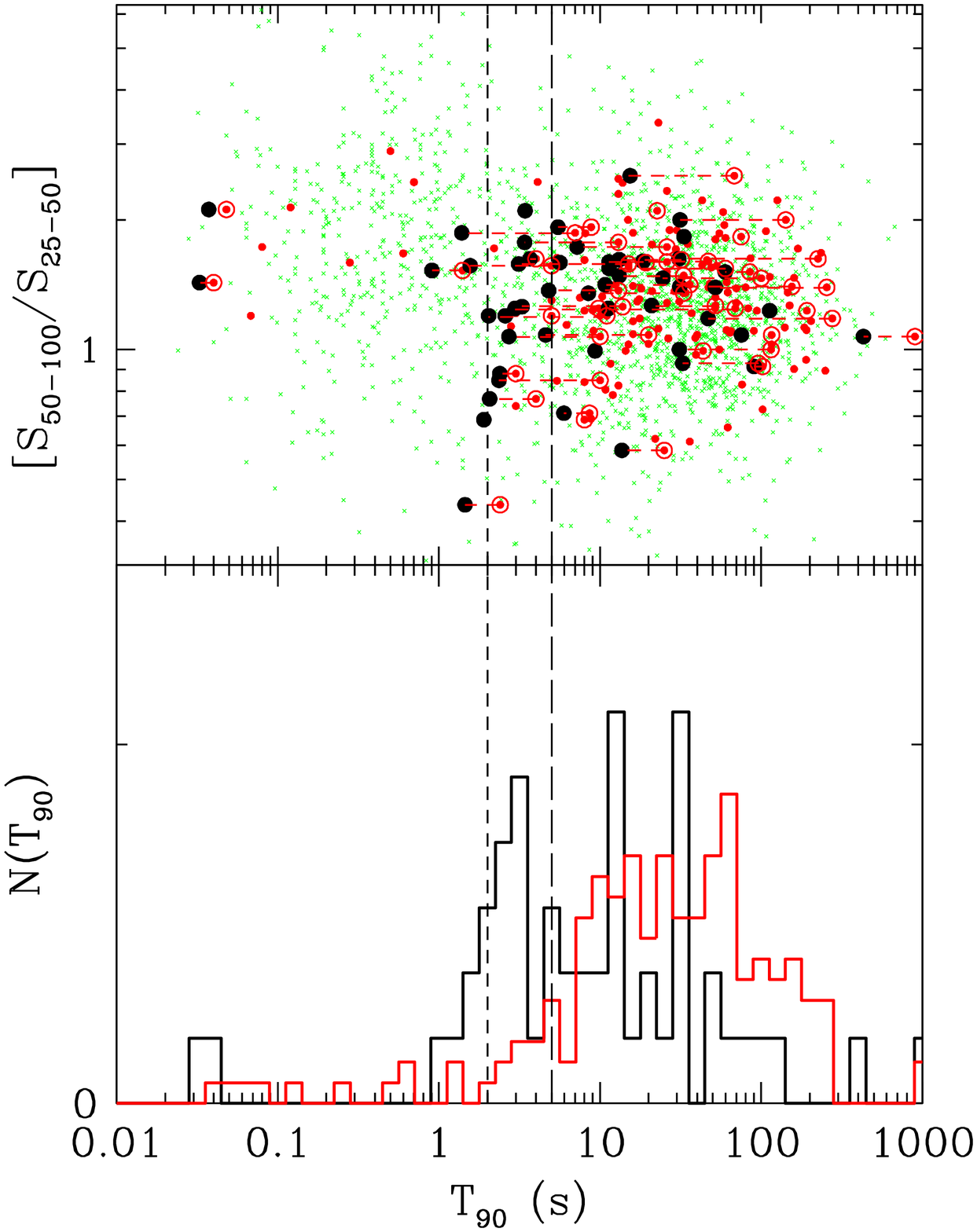}}}
\end{center}
\caption{
The hardness duration plot for GRBs detected by BATSE (green crosses)
showing the short-hard (upper left) and long-soft (lower right)
bursts. Overplotted are bursts observed by {\it Swift}, the small red
circles represent all {\it Swift} bursts while the larger circles are
those with redshifts. We have transformed the measured fluences into
the BATSE passband assuming the best fitting spectral parameters
reported in the GCN.
black are the locations of the bursts {\it de-redshifted}, (i.e. as
they would appear at zero redshift). As can been seen the corrections
for the short bursts are generally small, while much larger
corrections are necessary for the long bursts. Thus, the true duration
distributions of the two classes of bursts have a substantially larger
overlap than the observed ones. The two vertical lines show the
cannonical 2 second divide between long and short bursts, and the 5 second
divide suggested by Donaghy et al. (2006). We do not attempt to compensate for the
difference sensitivity of BAT compared to BATSE, and assume that the
power-law behaviour in the spectra extends across a broad enough range
that the redshift does not change the hardness ratio. In practice this
is likely often not the case, and thus bursts appear softer at high
redshift than they would at a lower redshift.  As such this plot
should be considered illustrative, rather than definitive. The lower panel shows
a histogram of the durations of all {\it Swift} bursts (red line) and the de-redshifted
durations of those with known redshifts.}
\label{f1}
\end{figure}

This effect may be present to differing degrees in the observed sample
of GRBs since it depends on the redshift distribution of the two populations.
If long and short duration bursts have the same redshift distribution then the
effects of redshift can essentially be neglected. However,
if short burst redshifts are typically low
(e.g. Nakar et al. 2006),  compared to long duration 
GRBs (Jakobsson et al. 2006a), then this effect may be 
significant. Therefore it is useful to consider
how the GRB duration distribution may appear in the {\em rest-frame} 
of the bursts.  This is shown in Figure 5, where the location of {\it Swift} 
bursts are shown as they were observed and as they would appear at zero
redshift\footnote{Note that given the spectral range of the BAT the precise measurements
of break energies in the prompt emission are not typically made, and therefore
the bursts have been shifted assuming a single power-law}. 
The duration of the bursts in their rest frame has been shortened by 
a factor of $(1+z)$, however, at lower luminosity distance it is possible to
track the emission out to late times, and therefore a factor of $(1+z)^{0.6}$ has
been suggested by Donaghy et al. (2006), by not attempting to 
consider this relation here we essentially can examine a worst case
scenario for the long-short duration overlap. 

As can be seen in Figure 5, there are several long duration bursts, which, 
if viewed at lower redshift may have been classified as short duration
bursts, these particular examples are GRB 050922C, GRB/XRF 050416,
GRB/XRF 050406 and GRB 051016. 
Additionally prior to {\it Swift}, GRB 000301C at $z=2.04$ exhibited a duration
of $\sim 2$s (Jensen et al. 2000) and GRB 040924 at $z=0.857$ had $t_{90} 
\sim 2 s$, although notably also exhibited a supernova (Soderberg et al. 2005). 

These bursts demonstrate overlap in the burst population culled purely
on duration. It may therefore be the case that we have observed a
larger population of short duration bursts that is currently
understood since the bursts may have been at higher redshifts, are
not distinct energetically from long duration bursts, and lie at too
great a distance for supernova searches to be attempted.

\subsection{Spectral lag - another method for short -- long discrimination?}
The determination of which population a given burst belongs to remains
of great importance and is an especially difficult task for bursts
with moderately short durations ($t_{90} \sim 3-8 s$) where there is
most overlap between the two classes.  In the case of GRB 060912A, the
measured spectral lag of $190^{+28}_{-40}$ms between the 5-25 and
50-100 keV bands and $83^{+43}_{-43}$ ms between the 25-50 and 100-350
bands placed the burst in long burst category, and, at early times
following the burst 
suggested,
apparently correctly, that the burst should be classified as long (Parsons et al. 2006).
This suggests, that {\em in cases where the duration of the burst is
close to 2s} spectral lag may be a good means of discriminating
between populations, with the advantage that it can be done rapidly,
and based purely on the prompt emission properties (Norris \& Bonnell
2006).

However, there may still be problems with the lag analysis in
understanding the properties of the burst. Specifically, the primary
aim of distinguishing the physical origin of the bursts. For example,
while long bursts apparently follow a reasonably constrained lag-peak
luminosity relationship (Gehrels et al. 2006), several of the GRBs
most convincingly associated with supernovae (e.g. GRB 980425, 031203)
lie off this empirical relation. Further GRB 060614 despite a duration
of $>$ 100s also shows a low lag measurement. Does this imply that it
really belongs to the same progenitor class as the short duration
bursts? If so it would stretch the plausibility of compact binary
mergers, since NS-NS mergers should be over very rapidly (e.g Rosswog
et al. 2002), while even for NS-BH mergers, which can show some
extended emission due periods of mass transfer during the inspiral
(Rosswog et al. 2004; Davies et al. 2005) may struggle to reproduce $>$100s of high
energy activity.

\section{Conclusions}
We have presented observations of GRB 060912A and its host galaxy.
Although the burst occurred very close to a bright elliptical galaxy,
lying within a group or cluster of galaxies at $z=0.0936$ (the coincidence
with the elliptical galaxy randomly has
a probability of only $7 \times 10^{-3}$) we conclude that it more likely
originated in an actively star forming galaxy at $z=0.937$.  The
properties of the burst, and of its host galaxy at this redshift
strongly suggest that GRB 060912A was a long duration burst, despite
several lines initially pointing to a short population burst. This
burst provides several important pointers for distinguishing between
the long and short GRB population:
\begin{enumerate}
\item
 {\em until} the properties of short burst hosts are better
constrained by a larger sample of bursts we should be
cautious is the assignation of host galaxies to nearby bright
galaxies, and putative host should not be
used as a strong indicator of burst type; 
\item
the difficulties in unambiguously 
identifying host galaxies, especially from X-ray only positions make
the use of other proxies which rely on distance (e.g. supernova presence or total energy)
unreliable.
\item If the true redshift distribution of short bursts is skewed to
lower redshifts than for the long duration population then the overlap in
rest-frame durations is larger, further blurring the distinction between 
long and short GRBs. 
\end{enumerate}

Notably however, a lag measurement demonstrated a positive lag and
indicated accurately the true nature of this burst. Indeed, although
lag measurements are also affected by cosmological time dilation the
de-redshifted lag-luminosity relations are more robust (though not
perfect) and may still allow distinction between different progenitor
types.

Finally, it should be noted that the issue of distinguishing between
different burst populations is not just one of curiosity for pursuing
followup observations, but is, in itself vital to constraining the
nature and range of their progenitors.  {\it Swift} observations are
demonstrating that GRB populations are markedly more diverse than had
previously been anticipated. This complicates the distinctions between
them and implies that the long-short divide does not adequately
describe all GRB populations. Attempting to simply place a burst in
one of only two categories may inhibit, rather than enhance our
knowledge of these still enigmatic transients.  Indeed, it is cases of
uncertainty (e.g. GRBs 060505, 060614 and 060912A) which may offer the
best means of understanding the observed GRB populations.

\section*{Acknowledgements}
We thank the referee for a constructive report which improved the paper. 
We also thank Sylvio Klose for comments on the manuscript.
AJL, CH, ER and NRT thank PPARC for support through fellowships
and rolling grant allocations. UK {\it Swift} research is also supported by
PPARC. PJ acknowledges support by a Marie Curie Intra-European
Fellowship within the 6th European Community Framework
Program under contract number MEIF-CT-2006-042001. 
The activity  of JG is  supported by  the Spanish  research programmes
    ESP2005-07714-C03-03 and AYA2004-01515. POB thanks the
    University of Leicester for study leave. RC acknowledges the support
    of a studentship from the University of Hertfordshire.


\begin{thebibliography}{99}

\bibitem[Barthelmy et al.(2005)]{2005Natur.438..994B} Barthelmy, S.~D., et 
al.\ 2005, Nature, 438, 994 


\bibitem[Berger et al.(2006)]{2006astro.ph.11128B} Berger, E., et al.\ 
2006a, ArXiv Astrophysics e-prints, arXiv:astro-ph/0611128 


\bibitem[Berger et al.(2006)]{2006astro.ph..8498B} Berger, E., Shin, M.~-., 
Mulchaey, J.~S., \& Jeltema, T.~E.\ 2006b, ArXiv Astrophysics e-prints, 
arXiv:astro-ph/0608498 



\bibitem[Berger(2006)]{2006GCN..5568....1B} Berger, E.\ 2006, GRB 
Coordinates Network, 5568

\bibitem[Berger et al.(2005)]{2005Natur.438..988B} Berger, E., et al.\ 
2005, Nature, 438, 988 




\bibitem[Bloom et al.(1998)]{1998ApJ...507L..25B} Bloom, J.~S., Djorgovski, 
S.~G., Kulkarni, S.~R., \& Frail, D.~A.\ 1998, ApJL, 507, L25 


\bibitem[Bloom et al.(2002)]{2002AJ....123.1111B} Bloom, J.~S., Kulkarni, 
S.~R., \& Djorgovski, S.~G.\ 2002, AJ, 123, 1111 


\bibitem[Bloom et al.(2003)]{2003ApJ...594..674B} Bloom, J.~S., Frail, 
D.~A., \& Kulkarni, S.~R.\ 2003, ApJ, 594, 674 


\bibitem[Bloom et al.(2006)]{2006ApJ...638..354B} Bloom, J.~S., et al.\ 
2006b, ApJ, 638, 354 


\bibitem[Bloom et al.(2007)]{2007ApJ...654..878B} Bloom, J.~S., et al.\ 
2007, ApJ, 654, 878 




\bibitem[Brown \& Hurkett(2006)]{2006GCN..5565....1B} Brown, P.~J., \& 
Hurkett, C.~P.\ 2006, GRB Coordinates Network, 5565, 1 


\bibitem[Butler(2007)]{2007AJ....133.1027B} Butler, N.~R.\ 2007, AJ, 133, 
1027 


\bibitem[Christensen et al.(2004)]{2004A&A...425..913C} Christensen, L., 
Hjorth, J., \& Gorosabel, J.\ 2004, A\&A, 425, 913 


\bibitem[Conselice et al.(2005)]{2005ApJ...633...29C} Conselice, C.~J., et 
al.\ 2005, ApJ, 633, 29 


\bibitem[Brown et al.(2006)]{2006GCN..5562....1H} Brown, P.  
\& Hurkett, C.\ 2006, GRB Coordinates Network, 5565


\bibitem[Della Valle et al.(2006)]{2006Natur.444.1050D} Della Valle, M., et 
al.\ 2006, Nature, 444, 1050 

\bibitem[Donaghy et al.(2006)]{2006astro.ph..5570D} Donaghy, T.~Q., et al.\ 
2006, ArXiv Astrophysics e-prints, arXiv:astro-ph/0605570 


\bibitem[Davies et al.(2005)]{2005MNRAS.356...54D} Davies, M.~B., Levan, 
A.~J., \& King, A.~R.\ 2005, MNRAS, 356, 54 


\bibitem[Ferrero et al.(2006)]{2006astro.ph.10255F} Ferrero, P., et al.\ 
2006, ArXiv Astrophysics e-prints, arXiv:astro-ph/0610255 


\bibitem[Fox et al.(2005)]{2005Natur.437..845F} Fox, D.~B., et al.\ 2005, 
Nature, 437, 845 


\bibitem[Fruchter et al.(2006)]{2006Natur.441..463F} Fruchter, A.~S., et 
al.\ 2006, Nature, 441, 463 


\bibitem[Fryer et al.(2006)]{2006ApJ...650.1028F} Fryer, C.~L., Young, 
P.~A., \& Hungerford, A.~L.\ 2006, ApJ, 650, 1028 



\bibitem[Fynbo et al.(2006)]{2006Natur.444.1047F} Fynbo, J.~P.~U., et al.\ 
2006, Nature, 444, 1047 


\bibitem[Gal-Yam et al.(2006)]{2006Natur.444.1053G} Gal-Yam, A., et al.\ 
2006, Nature, 444, 1053 






\bibitem[Gehrels et al.(2005)]{2005Natur.437..851G} Gehrels, N., et al.\ 
2005, Nature, 437, 851 


\bibitem[Gehrels et al.(2006)]{2006Natur.444.1044G} Gehrels, N., et al.\ 
2006, Nature, 444, 1044 


\bibitem[Glazebrook et al.(1995)]{1995MNRAS.273..157G} Glazebrook, K., 
Ellis, R., Colless, M., Broadhurst, T., Allington-Smith, J., \& Tanvir, N.\ 
1995, MNRAS, 273, 157 

\bibitem[Heise et al.(2001)]{2001grba.conf...16H} Heise, J., in't Zand, J., 
Kippen, R.~M., \& Woods, P.~M.\ 2001, Gamma-ray Bursts in the Afterglow 
Era, 16 



\bibitem[Hjorth et al.(2003)]{2003Natur.423..847H} Hjorth, J., et al.\ 
2003, Nature, 423, 847 


\bibitem[Hjorth et al.(2005)]{2005Natur.437..859H} Hjorth, J., et al.\ 
2005a, Nature, 437, 859 


\bibitem[Hjorth et al.(2005)]{2005ApJ...630L.117H} Hjorth, J., et al.\ 
2005b, ApJL, 630, L117 


\bibitem[Hogg et al.(1997)]{1997MNRAS.288..404H} Hogg, D.~W., Pahre, M.~A., 
McCarthy, J.~K., Cohen, J.~G., Blandford, R., Smail, I., \& Soifer, B.~T.\ 
1997, MNRAS, 288, 404 


\bibitem[Hurkett et al.(2006)]{2006GCN..5558....1H} Hurkett, C.~P., et al.\ 
2006a, GRB Coordinates Network, 5558


\bibitem[Hurkett et al.(2006)]{2006GCN..5562....1H} Hurkett, C.~P., Page, 
K.~L., \& Rol, E.\ 2006b, GRB Coordinates Network, 5562


\bibitem[Jakobsson et al.(2006)]{2006A&A...447..897J} Jakobsson, P., et 
al.\ 2006a, A\&A, 447, 897 

\bibitem[Jakobsson et al.(2006)]{2006GCN..5558....1H} Jakobsson, P., et al.\ 
2006b, GRB Coordinates Network, 5617


\bibitem[Jensen et al.(2001)]{2001A&A...370..909J} Jensen, B.~L., et al.\ 
2001, A\&A, 370, 909 


\bibitem[Kennicutt(1998)]{1998ARA&A..36..189K} Kennicutt, R.~C., Jr.\ 1998, 
ARA\&A, 36, 189 



\bibitem[Kouveliotou et al.(1993)]{1993ApJ...413L.101K} Kouveliotou, C., 
Meegan, C.~A., Fishman, G.~J., Bhat, N.~P., Briggs, M.~S., Koshut, T.~M., 
Paciesas, W.~S., \& Pendleton, G.~N.\ 1993, ApJL, 413, L101 

\bibitem[Le Floc'h et al.(2003)]{2003A&A...400..499L} Le Floc'h, E., et 
al.\ 2003, A\&A, 400, 499 


\bibitem[Levan et al.(2006)]{2006GCN..5558....1H} Levan, A.~J., et al.\ 
2006a, GRB Coordinates Network, 5573

\bibitem[Levan et al.(2006)]{2006ApJ...648L...9L} Levan, A.~J., et al.\ 
2006b, ApJL, 648, L9 

\bibitem[Levan et al.(2006)]{2006ApJ...647..471L} Levan, A., et al.\ 2006c, 
ApJ, 647, 471 


\bibitem[Nakar et al.(2006)]{2006ApJ...650..281N} Nakar, E., Gal-Yam, A., 
\& Fox, D.~B.\ 2006, ApJ, 650, 281 




\bibitem[Norris \& Bonnell(2006)]{2006ApJ...643..266N} Norris, J.~P., \& 
Bonnell, J.~T.\ 2006, ApJ, 643, 266 


\bibitem[Ofek et al.(2007)]{2007astro.ph..3192O} Ofek, E.~O., et al.\ 2007, 
ArXiv Astrophysics e-prints, arXiv:astro-ph/0703192 


\bibitem[Page et al.(2006)]{2006ApJ...637L..13P} Page, K.~L., et al.\ 2006, 
ApJL, 637, L13 

\bibitem[Parsons et al.(2006)]{2006GCN..5558....1H} Parsons, A., et al.\ 
2006, GRB Coordinates Network, 5561


\bibitem[Pedersen et al.(2005)]{2005ApJ...634L..17P} Pedersen, K., et al.\ 
2005, ApJ, 634, L17 



\bibitem[Rosswog et al.(2004)]{2004MNRAS.351.1121R} Rosswog, S., Speith, 
R., \& Wynn, G.~A.\ 2004, MNRAS, 351, 1121 



\bibitem[Soderberg et al.(2006)]{2006ApJ...636..391S} Soderberg, A.~M., et 
al.\ 2006, ApJ, 636, 391 




\bibitem[Vreeswijk et al.(2001)]{2001ApJ...546..672V} Vreeswijk, P.~M., et 
al.\ 2001, ApJ, 546, 672 


\bibitem[Vreeswijk et al.(2006)]{2006A&A...447..145V} Vreeswijk, P.~M., et 
al.\ 2006, A\&A, 447, 145 


\bibitem[Wainwright et al.(2005)]{2005astro.ph..8061W} Wainwright, C., 
Berger, E., \& Penprase, B.~E.\ 2005, ArXiv Astrophysics e-prints, 
arXiv:astro-ph/0508061 


\bibitem[Zeh et al.(2004)]{2004ApJ...609..952Z} Zeh, A., Klose, S., \& 
Hartmann, D.~H.\ 2004, ApJ, 609, 952 


\bibitem[Zhang et al.(2007)]{2007ApJ...655L..25Z} Zhang, B., Zhang, B.-B., 
Liang, E.-W., Gehrels, N., Burrows, D.~N., \& M{\'e}sz{\'a}ros, P.\ 2007, 
ApJL, 655, L25 


\bibitem[Zheng \& Ramirez-Ruiz(2006)]{2006astro.ph..1622Z} Zheng, Z., \& 
Ramirez-Ruiz, E.\ 2006, ArXiv Astrophysics e-prints, arXiv:astro-ph/0601622 







\end{thebibliography}
\end{document}